\documentclass[aps,prb,preprint,showpacs,superscriptaddress,floatfix]{revtex4-1}
\usepackage{graphics}
\usepackage{amsmath}
\usepackage{amssymb}
\usepackage{calc}
\usepackage{epsfig}
\usepackage{color}

\begin{document}

\title{ Magneto-optical control of F\"{o}rster energy transfer }

\author{R. Vincent}
\email{Corresponding author: remi.vincent@espci.fr}
\affiliation{Institut Langevin, ESPCI ParisTech, CNRS, 10 rue Vauquelin,
75231 Paris Cedex 05, France}

\author{R. Carminati}
\affiliation{Institut Langevin, ESPCI ParisTech, CNRS, 10 rue Vauquelin,
75231 Paris Cedex 05, France}
%

%

\date{\today}

%
%

\begin{abstract}

We introduce a general framework to study dipole-dipole energy transfer between an emitter and an absorber in a nanostructured
environment. The theory allows us to address F\"orster Resonant Energy Transfer (FRET) between a donor and an acceptor in the
presence of a nanoparticle with an anisotropic electromagnetic response.
In the particular case of a magneto-optical anisotropy, we compute
the generalized FRET rate and discuss the orders of magnitude. The distance dependence, the FRET efficiency
and the sensitivity to the orientation of the transition dipoles orientation differ from standard FRET and can be
controlled using the static magnetic field as an external parameter.
{\bf Keyword}: Fluorescence; FRET; surface plasmon; F\"orster radius; Magneto-optics; Nanoparticle; Quenching.
 \end{abstract}

\maketitle

\section{ Introduction}

Energy transfer between a molecule in an excited state
(donor) and a molecule in the ground state (acceptor) underlies many significant
photophysical and photochemical processes, from photosynthesis
to fluorescence probing of biological systems. It is also of interest in nanophotonics where efficient transfer of optical excitations on subwavelength scales is a key issue. Depending
on the separation between the donor (D) and the
acceptor (A), the process can be described accurately by
various theories accounting for the electromagnetic interaction
between the two species. For a D-A distance range
on the order of 2-10 nm, which is relevant for photochemical studies and nanophotonics, the well established
F\"{o}rster theory \cite{Forster59} based on quasi-static dipole-dipole
interaction has been very successful. It shows that while FRET is a very useful process
which can be used, for example, as a ruler for spectroscopic
measurements\cite{Stryer78}, it is a rather weak process which goes
down as the inverse sixth power $R^{-6}$ of the D-A separation.
In fact, one can introduce a length scale
known as the F\"{o}rster distance $R_0$ at which FRET is 50\%
efficient and it is found that $R_0$ is on the order of a few nanometers in most
practical situations.
For even smaller distances, Dexter \cite{Dexter53} recognized that
electronic exchange and multipolar interactions become important
and a full quantum mechanical treatment must be
implemented. On the other hand, in the large distance regime (non-negligible compared to the
wavelength), full electrodynamics is needed to account for retardation effects.
In this work, we will focus on the so-called F\"{o}rster resonance energy transfer
(FRET) when D and A are located in the vicinity of a
nanoparticle. In this three-body configuration, we will extend the FRET formalism and show that the presence of the external nanostructure introduces interesting degrees of freedom.
In the case of a nanoparticle with an anisotropic dielectric response (e.g., a nanoparticle made
of a ferromagnetic material exhibiting a magneto-optical response), the distance dependence, the orientation dependence and the strength of the FRET efficiency can be changed substantially.
In the case of a magneto-optical anisotropy,
it can in principle be controlled using the static magnetic field as an external control parameter.

\section{Generalized FRET Theory}

In this section, we introduce a generalized formalism to compute the FRET rate
that allows us to deal with a D-A system in interaction with
a nanostructured environment.
This formalism includes as a particular case the standard
F\"{o}rster  theory.
\begin{figure}
\centerline{\resizebox{09.cm}{!}{\includegraphics{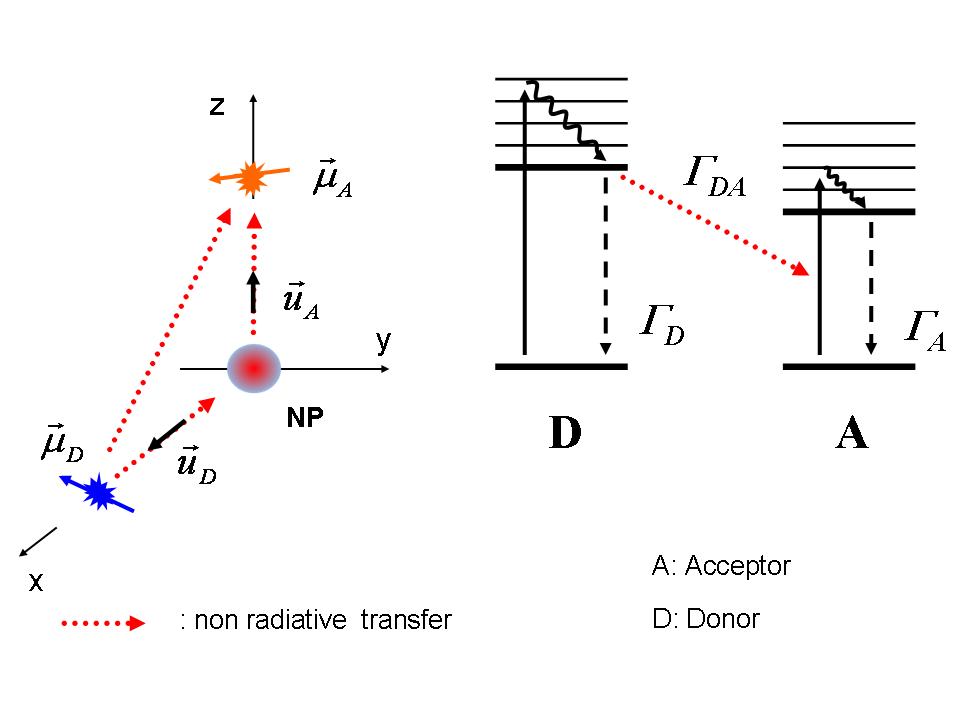}}}
\caption{\label{slab}Left panel: Schematic configuration of the D-A system in the presence of a nanoparticle.
The different channels for energy transfer (direct or indirect) are indicated by dotted arrows.
When the transition dipoles are orthogonal, the direct F\"orster transfer is disabled.
Right panel: Energy-level diagram of the donor and acceptor molecules.  }
\end{figure}
Let us consider a donor (emitter) and an acceptor (absorber) with arbitrary locations
and orientations of transition dipoles, in the vicinity of a nanostructure, as illustrated in Fig. 1 (a single nanoparticle
is represented for the sake of illustration, but the theory presented in this section is not restricted to this geometry).
We denote  by $(\boldsymbol{ r}_A,\boldsymbol{\mu}_A) $
and by $(\boldsymbol{r}_D, \boldsymbol{\mu}_D)  $ the
position and the direction of the transition dipole of the acceptor and donor, respectively.
For electric-dipole transitions and in the weak-coupling regime, the normalized FRET rate $\Gamma_{DA}/\Gamma_0$ can be calculated from the electric Green function which describes the electromagnetic response of the environment. It takes the form (the full derivation is given in Appendix A):
\begin{eqnarray}
\frac{\Gamma_{DA}}{\Gamma_0}=18\pi \epsilon_0^2 c^4 \int_0^{\infty} \frac{\sigma_A(\omega)f_D(\omega)M(\omega)}{\omega^4}\text{d} \omega.
\end{eqnarray}
In this expression, $\Gamma_{DA}$ is the energy transfer rate from donor to acceptor, $\Gamma_0$ the decay rate of the donor in free space and $\omega$ the emission frequency. $f_D(\omega) $ is the normalized emission spectrum of the donor, and the function $M$ is defined by
\begin{eqnarray}
M(\omega)= |\boldsymbol{\mu}_A.\boldsymbol{G}(\boldsymbol{r}_A,\boldsymbol{ r}_D,\omega).\boldsymbol{\mu}_D|^2
\end{eqnarray}
where $\bf{G}$ is the electric dyadic Green function that describes the environment of the donor and acceptor. It is defined as follows: For a point electric dipole ${\bf p}$ located at position ${\bf r}^\prime$ and oscillating at frequency $\omega$, the electric field radiated at position ${\bf r}$ reads ${\bf E}({\bf r})=\mu_0\omega^2{\bf G}({\bf r},{\bf r}^\prime,\omega){\bf p}$. The expression of the FRET rate given in Eqs (1-2) is very general, and can be applied to an arbitrary geometry, provided that the Green dyadic is known. Thus Eqs (1-2) establish the basis for a generalized FRET theory. In particular, it shows that the FRET signal, as any fluorescence signal, is strongly dependent on the environment~\cite{Novotny}.

\subsection{FRET rate in free space}

To recover the standard FRET formalism corresponding to a D-A couple in free space, one can rewrite Eq. (1) as follows
\begin{eqnarray}
\frac{\Gamma_{DA}}{\Gamma_0}= \frac{9c^4}{8\pi R^6}\int_0^{\infty} \frac{\sigma_A(\omega)f_D(\omega)T(\omega,R)}{\omega^4}\text{d} \omega
\end{eqnarray}
with
\begin{eqnarray}
T(\omega,R)= 16\pi^2\epsilon_0^2R^6|\boldsymbol{\mu}_A .\boldsymbol{G}(\boldsymbol{r}_A,\boldsymbol{ r}_D,\omega). \boldsymbol{\mu}_D|^2
\end{eqnarray}
and simply replace $\bf{G}$ by the dyadic Green function $\bf{G}_0$ of free space. In the quasi-static limit $\boldsymbol{G}_0(\boldsymbol{r},\boldsymbol{r}',\omega)=1/(4\pi \epsilon_0)(3\boldsymbol{ v}\boldsymbol{v}-\boldsymbol{1})/|\boldsymbol{r}-\boldsymbol{r}'|^3$, with $\boldsymbol{v}=(\boldsymbol{r}-\boldsymbol{r}')/|\boldsymbol{r}-\boldsymbol{r}'|$ the unit vector in the direction of ($\boldsymbol{r}-\boldsymbol{r}'$). This leads to the well-known expression of the standard FRET rate $\Gamma^0_{DA} $ \cite{Novotny,Forster59,Gersten84,Hua85}:
\begin{eqnarray}
\frac{\Gamma^0_{DA}}{\Gamma_0}= \left(\frac{R_0}{R}\right)^6
\end{eqnarray}
where $R=|\boldsymbol{r}_A-\boldsymbol{r}_D|$ is the distance between the acceptor and the donor, and where the F\"orster radius $R_0$ is readily identified as follows
\begin{eqnarray}
R_0^6= \frac{9c^4 \kappa^2}{8\pi }\int_0^{\infty} \frac{\sigma_A(\omega)f_D(\omega)}{\omega^4}\text{d} \omega .
\end{eqnarray}
In this expression, $\kappa=3( \boldsymbol{u} .\boldsymbol{\mu}_D )( \boldsymbol{u} .\boldsymbol{ \mu}_A )    -\boldsymbol{\mu}_A . \boldsymbol{\mu}_D  $
is the orientational factor,  with $\boldsymbol{ u}=(\boldsymbol{r}_D-\boldsymbol{r}_A )/| \boldsymbol{r}_D-\boldsymbol{r}_A |$ the unit vector along the axis
of the D-A couple. The orientational factor can take values from 0 (perpendicular transition dipoles) to 2 (parallel transition dipoles).

\subsection{FRET rate in the presence of a nanoparticle with an anisotropic dielectric response }

In principle, the presence of a nanostructure close to a D-A couple
will modify the emission and absorption by the transition dipoles. The modifications are accounted for by the dyadic Green function that describes
the electrodynamic response of the environment, through the function $M(\omega)$ entering Eq. (1).
This formalism leads to a very general treatment of the FRET transfer mediated by an external nanostructure, such as a nanoparticle with an
anisotropic response. It permits a study of
the influence of many parameters of practical relevance, such as the orientation of the transition dipoles, or the shape and material properties of the nanoparticle.

In the presence of the nanoparticle, described itself in the electric-dipole approximation, the full dyadic Green function reads
\begin{eqnarray}
\boldsymbol{G}(\boldsymbol{r},\boldsymbol{r}',\omega)=\boldsymbol{G}_0(\boldsymbol{r},\boldsymbol{ r}',\omega)+ \nonumber\\ \boldsymbol{G}_0(\boldsymbol{r},\boldsymbol{r}_p,\omega)\cdot \boldsymbol{ \alpha}(\omega)\epsilon_0 \cdot \boldsymbol{G}_0(\boldsymbol{r}_p,\boldsymbol{ r}',\omega)\label{eqGreen}
\end{eqnarray}
where  $\boldsymbol{r}_p $ is the center of the nanoparticle and $\boldsymbol{\alpha}(\omega)$ its polarizability tensor.
In the following we assume that the nanoparticle is located at the origin and set $\boldsymbol{r}_p =0$.
Equation (1) can be rewritten as follows
\begin{eqnarray}
\frac{\Gamma_{DA}}{\Gamma_0}=18\pi \epsilon_0^2 c^4 \int_0^{\infty} \frac{\sigma_A(\omega)f_D(\omega)M(\omega)}{\omega^4}\text{d} \omega
\end{eqnarray}
where the function $M$ is given by
\begin{eqnarray}
M(\omega)= \left |\boldsymbol{\mu}_A .\boldsymbol{G}(\boldsymbol{r}_A,\boldsymbol{ r}_D,\omega).\boldsymbol{\mu}_D \right |^2\nonumber\\
=\left |\boldsymbol{\mu}_A .\boldsymbol{G}_0(\boldsymbol{r}_A,\boldsymbol{ r}_D,\omega).\boldsymbol{\mu}_D+\boldsymbol{\mu}_A .\boldsymbol{G}_0(\boldsymbol{r}_A,0,\omega).\boldsymbol{\alpha}(\omega)\epsilon_0\boldsymbol{G}_0(0,\boldsymbol{r}_D
,\omega).\boldsymbol{\mu}_D \right |^2.
\end{eqnarray}
Using again the quasi-static limit for the free-space dyadic Green function $\boldsymbol{G}_0$, we obtain
\begin{eqnarray}
M(\omega)= \left |\frac{3( \boldsymbol{u} .\boldsymbol{\mu}_D )( \boldsymbol{u} .\boldsymbol{\mu}_A )    -\boldsymbol{ \mu}_A . \boldsymbol{\mu}_D}{4\pi \epsilon_0 R^3}+\boldsymbol{ \mu}_A .\frac{3\boldsymbol{ u}_A.\boldsymbol{u}_A -\boldsymbol{1}   }{4\pi \epsilon_0 R_{A}^3}.\boldsymbol{ \alpha}(\omega)\epsilon_0\frac{3\boldsymbol{u}_D.\boldsymbol{u}_D -\boldsymbol{1}   }{4\pi \epsilon_0 R_{D}^3}.\boldsymbol{ \mu}_D \right |^2\nonumber\\
=\left |\frac{\kappa}{4\pi \epsilon_0 R^3}+\frac{\boldsymbol{\alpha}_{\boldsymbol{\mu}_A \boldsymbol{\mu}_D } +\boldsymbol{\alpha}_{\boldsymbol{u}_A \boldsymbol{u}_D }9  (\boldsymbol{u}_{D} .\boldsymbol{\mu}_D )( \boldsymbol{u}_{A} .\boldsymbol{\mu}_A )-3((\boldsymbol{u}_{D} .\boldsymbol{ \mu}_D)\boldsymbol{\alpha}_{\boldsymbol{\mu}_A \boldsymbol{u}_D }+(\boldsymbol{u}_{A} .\boldsymbol{ \mu}_A)\boldsymbol{\alpha}_{\boldsymbol{u}_A \boldsymbol{\mu}_D }) }{(4\pi)^2 \epsilon_0 R_D^3R_A^3} \right |^2\nonumber\\
.
\end{eqnarray}
where $R_A=|\boldsymbol{r}_A|$ is the distance between the acceptor and the nanoparticle, $R_D=|\boldsymbol{r}_D|$ is the distance between the donor and the nanoparticle, and $\boldsymbol{u}_A=\boldsymbol{r}_A/|\boldsymbol{r}_A|$, $\boldsymbol{u}_D=\boldsymbol{r}_D/|\boldsymbol{r}_D|$.
This expression clearly shows the contribution of the different non-radiative energy transfer channels: The direct (standard) F\"orster transfer $\Gamma^0_{DA}$, the energy transfer mediated by the nanoparticle $\Gamma^{NP}_{DA}$, and an interference term accounting for the phase shift between the two channels.
In the following, we will focus on the role of the nanoparticle and compute the FRET rate $\Gamma^{NP}_{DA}$.
In the spirit of the F\"orster radius, we introduce the distance $R_{p}$ such that
\begin{eqnarray}
R_{p}^6= \frac{ \int_0^{\infty} \displaystyle \frac{\sigma_A(\omega)f_D(\omega)|\boldsymbol{\alpha}_{\boldsymbol{\mu}_A \boldsymbol{\mu}_D } +\boldsymbol{ \alpha}_{\boldsymbol{u}_A \boldsymbol{u}_D }9  (\boldsymbol{u}_{D} .\boldsymbol{\mu}_D )( \boldsymbol{u}_{A} .\boldsymbol{\mu}_A )-3((\boldsymbol{u}_{D} .\boldsymbol{\mu}_D)\boldsymbol{ \alpha}_{\boldsymbol{\mu}_A \boldsymbol{u}_D }+(\boldsymbol{u}_{A} .\boldsymbol{\mu}_A)\boldsymbol{ \alpha}_{\boldsymbol{u}_A \boldsymbol{\mu}_D })|^2}{\omega^4}\text{d} \omega} { (4\pi)^2 \int_0^{\infty} \displaystyle \frac{\sigma_A(\omega)f_D(\omega)}{\omega^4}\text{d} \omega } .
\end{eqnarray}
This new length scale $R_{p}$ will be denoted by {\it polarization coupling radius}. Using this quantity, the FRET rate mediated by the nanoparticle can be
rewritten as
\begin{eqnarray}
\frac{\Gamma^{NP}_{DA}}{\Gamma_0}= \left(\frac{R_0(\kappa=1)}{R_A}\right)^6 \left(\frac{R_{p}}{R_D}\right)^6 .
\end{eqnarray}
This compact expression is convenient for the analysis of the FRET rate mediated by a nanoparticle, and its derivation is a key step in the present paper.
An important result is that the distance dependence differs from that of standard (free space) FRET.
Moreover, the polarization coupling radius $R_p$ allows us to compare the indirect FRET rate (i.e., mediated by the nanoparticle) and
the standard free-space FRET rate.

For the sake of illustration, let us consider the situation in which the three bodies are aligned (see Fig. 2(a)), with $R_D=R_A=2R_{NP}$ and $R=4R_{NP}$ ($\kappa=2$).
In this case, we obtain
\begin{eqnarray}
\frac{\Gamma^{NP}_{DA}}{\Gamma^0_{DA}}= \left(\frac{R_0(\kappa=1)}{R_0(\kappa=2)}\right)^6\left(\frac{R_{p}}{R_{NP}}\right)^6 = \frac{1}{4} \left(\frac{R_{p}}{R_{NP}}\right)^6 .
\end{eqnarray}
This simple expression shows that the ratio $R_p/R_{NP}$ is the crucial parameter that describes the influence of the nanoparticle
on the FRET rate. For $R_p>R_{NP}$, the nanoparticle enhances the FRET transfer, while for $R_p<<R_{NP}$, the FRET becomes exclusively driven by the direct transfer.
In order to get insight into the meaning of the polarization coupling radius, let us assume that the polarizability of the nanoparticle $\boldsymbol{ \alpha}(\omega)$ varies smoothly on the frequency range of the spectral overlap between $\sigma_A(\omega)$ and $f_D(\omega)$. We can rewrite Eq. (11) as follows
\begin{eqnarray}
R_{p}^3=\frac{ |\boldsymbol{\alpha}_{\boldsymbol{\mu}_A \boldsymbol{\mu}_D } +\boldsymbol{\alpha}_{\boldsymbol{u}_A \boldsymbol{u}_D }9  (\boldsymbol{u}_{D} .\boldsymbol{\mu}_D )( \boldsymbol{u}_{A} .\boldsymbol{\mu}_A )-3((\boldsymbol{u}_{D} .\boldsymbol{\mu}_D)\boldsymbol{ \alpha}_{\boldsymbol{\mu}_A \boldsymbol{u}_D }+(\boldsymbol{u}_{A} .\boldsymbol{\mu}_A)\boldsymbol{ \alpha}_{\boldsymbol{u}_A \boldsymbol{\mu}_D })|}{4\pi} .
\end{eqnarray}
In this condition, the polarization coupling radius $R_{p}$ depends only on the polarizability tensor of the nanoparticle.
Such a framework allows us to put forward explicitly the relevant length scales involved in the FRET mechanism mediated by the nanoparticle, i.e.,  the F\"orster radius $R_0$ and the polarization coupling radius $R_{p}$. In the following we use this formalism to calculate explicitly
the FRET rate of a D-A couple interacting with a spherical nanoparticle exhibiting a magneto-optical response or a purely metallic response.

\begin{figure}
\centerline{\resizebox{09.cm}{!}{\includegraphics{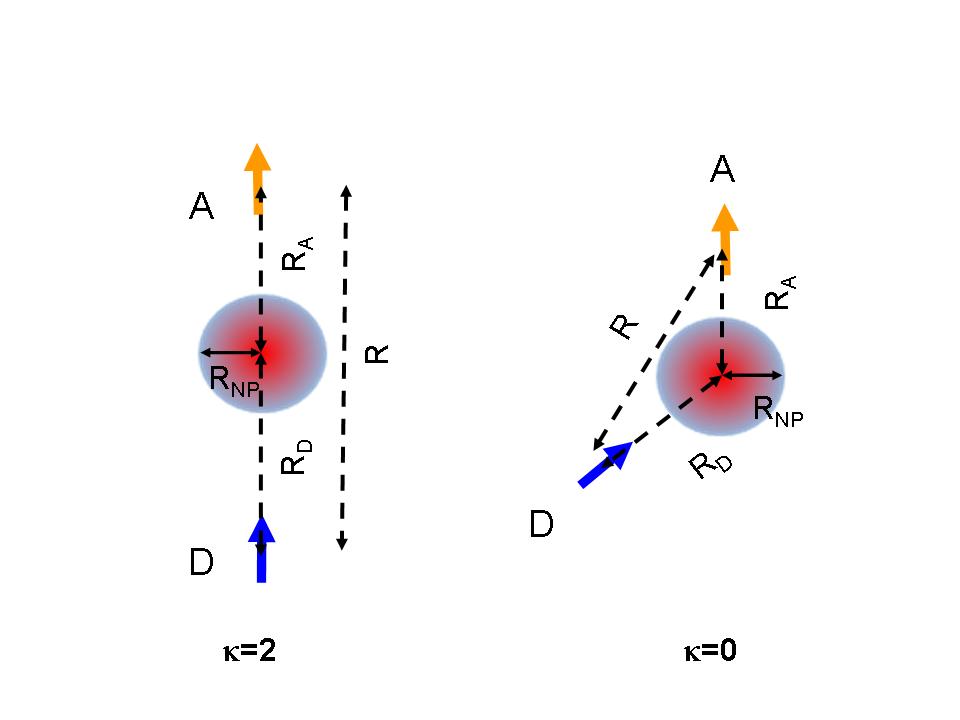}}}
\caption{\label{slab}Two canonical configurations studied in the present work. (a) Left panel: Aligned configuration.
(b) Right panel: Orthogonal configuration.  }
\end{figure}

\subsection{Spherical magneto-optical nanoparticle}

Ferromagnetic materials exhibit magnetic anisotropy that can be controlled by an external static magnetic field.
At saturation, their magnetization influences the dielectric function, that exhibits the so-called magneto-optical response.
A spherical nanoparticle made of a material with a magneto-optical response can be described by an anisotropic electric polarizability with radiative corrections \cite{Saenz10}
\begin{eqnarray}
\boldsymbol{\alpha}(\omega)=(\boldsymbol{I}-i\frac{k^3}{6\pi}\boldsymbol{ \alpha}_0)^{-1}\boldsymbol{\alpha}_0
\end{eqnarray}
where $\boldsymbol{\alpha_0}(\omega)$ is the quasi-static polarizability. For magneto-optical materials (gyrotropic material) in the presence of a static magnetic field, it reads
\begin{eqnarray}
\boldsymbol{\alpha}_0(\omega)=3V\frac{\boldsymbol{\epsilon}-\boldsymbol{ I}}{\boldsymbol{\epsilon}+2\boldsymbol{I}}
\end{eqnarray}
where $V$ is the volume of the nanoparticle and $\boldsymbol{\epsilon}$ is the dielectric tensor, given by
\begin{eqnarray}
\boldsymbol{\epsilon}=\epsilon_I \boldsymbol{I}+iQ\boldsymbol{A}=
\begin{pmatrix}
\epsilon_I &-iQm_z&iQm_y \\
iQm_z &\epsilon_I& -iQm_x \\
-iQm_y& iQm_x & \epsilon_I
\end{pmatrix}.
\end{eqnarray}
In this expression $\boldsymbol{m}=(m_x,m_y,m_z)$ is the direction of the magnetization inside the particle,  $\boldsymbol{\epsilon}_I$ the diagonal part of the dielectric tensor and $Q$ is the magneto-optical
coefficient.
Let us stress that expression (15) of the polarizability is consistent
with energy conservation (or, equivalently, the optical
theorem)\cite{Draine88,Lakhtatia90}. In the case of a pure metal, the same expression of the polarizability holds, with an isotropic
dielectric function $\boldsymbol{\epsilon}(\omega)=\epsilon(\omega) \boldsymbol{I}$. In the present paper, we use
bulk dielectric functions (no finite-size effects), which is a reasonable approach when the sizes involved remain larger than a few nanometers \cite{FordWeber84}.

\section{ Discussion}

The formalism in the previous section has shown the crucial role of the polarization coupling radius $R_p$ on the FRET rate mediated
by a nanoparticle. In this section, using experimental data for the dielectric function of different metallic and magneto-optical materials \cite{Palik85}, we compute the ratio $R_{p}/R_{NP}$, where $R_{NP}$ is the radius of the nanoparticle, and study its dependence on different parameters of practical relevance:  The emission wavelength of the donor, the radius of the nanoparticle, and the material properties.

\subsection{FRET mediated by a metallic nanoparticle}

Noble metals are known to hold plasmon resonances that enhance, for example, the polarizability of nanoparticle.
Since the polarization coupling radius $R_p$ directly depends on the polarizability, one can expect a substantial influence
of the plasmon resonance on the FRET rate mediated by the nanoparticle. This is indeed what we observe in Fig. 3, in which
we have plotted the ratio $R_{p}/R_{NP}$ with $R_{NP}=10$ nm) versus the emission wavelength of the donor for gold and silver, that  are common materials
in studies of fluorescence enhancement or quenching. We have considered the aligned configuration, with $\boldsymbol{u}_{D} =\boldsymbol{\mu}_D$, $\boldsymbol{u}_{A} =\boldsymbol{\mu}_A$, and $ \boldsymbol{\mu}_A.\boldsymbol{\mu}_D=1$ [see Fig. 2(a) for a sketch of the aligned geometry].
The plasmon resonance is visible in both cases, leading to an enhancement of $R_p/R_{NP}$. For instance in the case of silver, one reaches $R_{p}/R_{NP} \simeq 3$; for gold, one has $R_{p}/R_{NP} \simeq 1.9$. In the particular conditions $R_D=R_A=2R_{NP}$ and $R=4R_{NP}$, corresponding to the validity of Eq. (13),
we obtain an enhancement factor $\Gamma^{NP}_{DA}/\Gamma^0_{DA}$ of the FRET rate on the order of $180$ for silver and  $10$ for gold. For a D-A couple working at plasmon resonance with these materials, we conclude that FRET is mainly driven by the nanoparticle. Incidentally, any change of the
dielectric property of the nanoparticle will be reflected in a modulation of the FRET rate. Modulation of the dielectric response can, be achieved, e.g., through the magneto-optical effect that we consider in the following.

\begin{figure}
\centerline{\resizebox{09.cm}{!}{\includegraphics{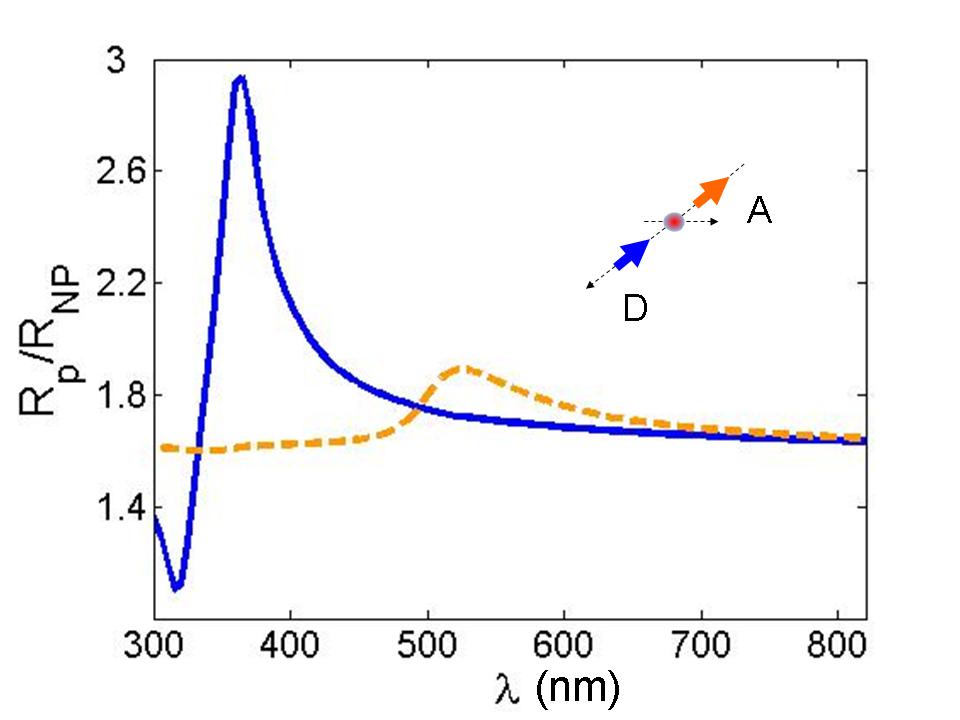}}}
\caption{\label{slab}Ratio  $R_{p}/R_{NP}$ for Silver (blue solid line), and Gold (gold dashed line) as a function of the excitation wavelength $\lambda$.  $R_{NP}=10 $ nm.  The configuration is such that the dipole are collinear and Acceptor, Donor and nanoparticle are aligned.   }
\end{figure}

\subsection{ Controlling FRET through the magneto-optical interaction}

A magneto-optical nanoparticle can be described using the polarizability in Eqs. (15-17), together with
experimental data for the dielectric function of well known magneto-optical materials \cite{Palik85}. In this section,
we compute the ratio $R_{p}/R_{NP}$ and study the influence of different parameters, such as donor emission wavelength, nanoparticle
material and size.

We show in Fig.4 the ratio $R_{p}/R_{NP}$ (with $R_{NP}=10$ nm) versus the emission wavelength of the donor for different materials that are known
to exhibit a magneto-optical response (Nickel, Iron and Cobalt), and in the same aligned configuration as in Fig. 3.  We observe a smoother behavior than in the case of noble metals (Fig. 3), since plasmon resonances are strongly damped by absorption in these magneto-optical materials.
For these materials, the enhancement factor $\Gamma^{NP}_{DA}/\Gamma^0_{DA}$ of the FRET rate is on the order of $5$, showing that in this case the FRET rate is also driven by the nanoparticle.
\begin{figure}
\centerline{\resizebox{09.cm}{!}{\includegraphics{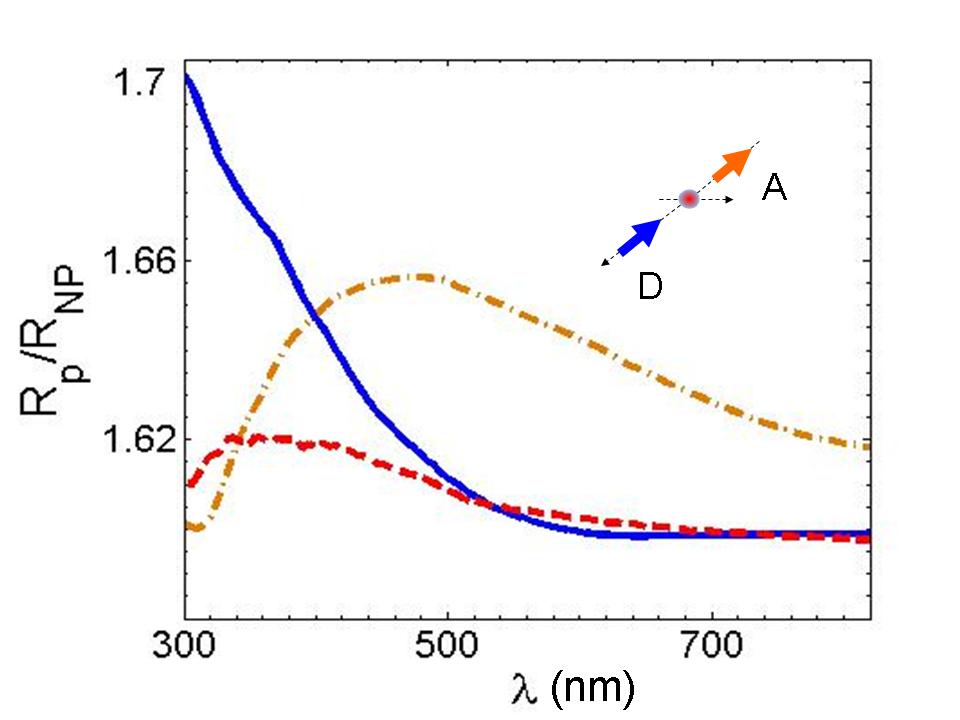}}}
\caption{\label{slab}Ratio  $R_{p}/R_{NP}$ for Iron (blue solid line), Nickel (gold dash-dotted line), and Cobalt (red dashed line) as a function of the emission wavelength $\lambda$ of the donor.  $R_{NP}=10 $ nm.  The configuration is illustrated in the inset, showing that the dipole are collinear and the couple Donor-Acceptor and nanoparticle are aligned. }
\end{figure}

Figure 5 shows a computation of the ratio $R_{p}/R_{NP}$ with the same materials as in Fig. 4, but in the case of an orthogonal configuration with $\boldsymbol{u}_{D} =\boldsymbol{\mu}_D$, $\boldsymbol{u}_{A} =\boldsymbol{\mu}_A$, and $ \boldsymbol{\mu}_A.\boldsymbol{\mu}_D=0$ [see Fig. 2(b) for a sketch of the orthogonal geometry]. The magnetization is orthogonal to the plane containing the D-A couple and the nanoparticle.
Let us stress that, in this configuration, the FRET rate vanishes in absence of an external static magnetic field due to the orthogonality of the donor and acceptor transition dipoles ($\kappa=0$). Although one observes that $R_{p}/R_{NP}$ remains smaller than one, the possibility of inducing a FRET rate driven only by the polarization anisotropy of the nanoparticle is an interesting result, showing the potential of magneto-optical nanoparticles for FRET.
On the one hand, the anisotropic response allows to couple molecules for which standard FRET gives a vanishing signal due to orientational mismatch.
On the other hand, the possibility of controlling the magneto-optical response with a static magnetic field as an external parameter could allow to tune or
modulate the FRET rate, which can be an advantage, e.g., to increase the sensitivity of the detection process.
\begin{figure}
\centerline{\resizebox{09.cm}{!}{\includegraphics{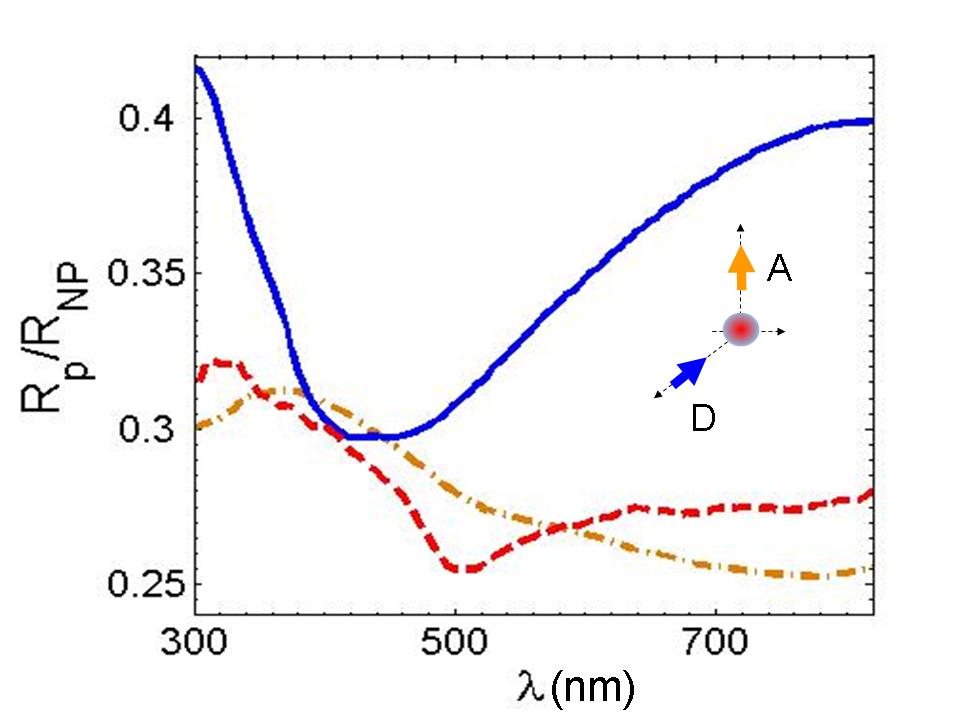}}}
\caption{\label{slab}Ratio  $R_{p}/R_{NP}$ for Iron (blue solid line), Nickel (yellow dash-dotted line), and Cobalt (red dashed line) as a function of the emission wavelength $\lambda$ of the donor  in the presence of an external magnetic field inducing a magnetization in the direction orthogonal of the plane containing the tree body D-A-NP.  $R_{NP}=10 $ nm. The configuration is illustrated in the inset ($(\boldsymbol{u}_{D} =\boldsymbol{\mu}_D )$, $( \boldsymbol{ u}_{A} =\boldsymbol{\mu}_A )$, and $ \boldsymbol{\mu}_A.\boldsymbol{\mu}_D=0$) }
\end{figure}

\subsection{Distance dependence of the FRET rate}

The $R^{-6}$ of free-space FRET (F\"orster's theory) is a well-known feature of non-radiative dipole-dipole-coupling. In the presence of an external
body, this distance dependence is obviously more involved, as shown by the general theory that we introduced.
In Eq. (12), we see a dependence on both the donor-nanoparticle distance ($R_D$) and the acceptor-nanoparticle distance ($R_A$).  In order to
 get insight into the dependence of the generalized FRET rate on the acceptor-nanoparticle distance only, we arbitrary fix the
 donor-nanoparticle distance as follows: $R_D=2R_{NP}$ with $R_{NP}\sim R_0 /2$ (note that the opposite choice can be made).
In this case, the dependence of the FRET rate on the acceptor-nanoparticle distance is given by
\begin{eqnarray}
\frac{\Gamma_{DA}^{NP}}{\Gamma_0}= \left(\frac{R_{p}/R_{NP}}{R_A/R_{NP}}\right)^6.
\end{eqnarray}
From this expression, we see that we recover an inverse sixth power dependence, as in standard FRET, but with the D-A distance replaced by the acceptor-nanoparticle distance. This simply reflects the mechanism underlying the FRET rate mediated by the nanoparticle channel, which can be seen as a series of non-radiative energy transfer from the donor to the nanoparticle followed by another transfer from the nanoparticle to the acceptor.
It is known that non-radiative energy transfer through dipole-dipole interaction between an emitter and a nanoparticle leads to the $R^{-6}$ distance dependence that is recovered here\cite{Carminati06,Seelig07,colasdes08}.
Another important output of
Eq. (18) is that from the knowledge of the ratio $R_{p}/R_{NP}$, we know exactly what the critical acceptor-nanoparticle distance that separates the quenching and enhancement regimes of FRET mediated by the nanoparticle.

\section{Conclusion}

We have solved the F\"orster energy transfer problem in a three-body configuration, involving two fluorophores close to a nanoparticle with an anisotropic dielectric response. Using a Green function formalism, we have shown that the angular contribution,
the distance behavior and the influence of the polarizability tensor of the nanoparticle can be identified and separated. The distance dependence is controlled by a new distance $R_p$ that depends of the polarization properties of the nanoparticle.
We have illustrated the formalism in the case of a magneto-optical nanoparticle for which the degree of anisotropy can be controlled by an external static magnetic field, and we have discussed potential application for FRET tuning and modulation.
Here, we have presented a proof of concept. Further work should focus on enhancing the (weak) magneto-optical FRET signal.
We have illustrated the formalism also  for the well known metallic nanoparticle, showing that this formalism could furnish insight in the understanding of the good quantities controlling this process.

\section{Appendix}
The expression of the normalized energy transfer can be obtained by treating the acceptor as classical harmonic damped dipole oscillating at frequency $\omega$. In this approach the normalized fluorescence energy transfer is equivalent to the normalized power emitted by the classical dipole. One writes $\Gamma_{DA}/\Gamma_0=P_{DA}/P_0$, where $P_{DA}$ is the power transmitted from donor to acceptor in the presence of the environment and $P_0$ is the power emitted by the donor in absence of acceptor and in free-space. $P_0$ can be written as
\begin{eqnarray}
P_0=\frac{\omega}{2}|\boldsymbol{p}_D|^2 Im[\boldsymbol{\mu}_D .\boldsymbol{G}_0(\boldsymbol{r}_D,\boldsymbol{ r}_D,\omega). \boldsymbol{\mu}_D ],
\end{eqnarray}
with $\boldsymbol{p}_D $, $\boldsymbol{p}_A $, the donor and acceptor dipole respectively.

The power absorbed by the acceptor can be written
\begin{eqnarray}
P_{DA}=-\frac{\omega}{2} Im[\boldsymbol{p}_A^*.\boldsymbol{G}(\boldsymbol{r}_A,\boldsymbol{ r}_D,\omega). \boldsymbol{p}_D ].
\end{eqnarray}
assuming that the acceptor is a polarizable molecule of fix direction ($\boldsymbol{\mu}_A$), with $ \boldsymbol{\alpha}_A=\alpha_A(\omega)\boldsymbol{\mu}_A\boldsymbol{\mu}_A$,  we may write
\begin{eqnarray}
\boldsymbol{p}_A=\boldsymbol{\alpha}_A\boldsymbol{E}_D(\boldsymbol{r}_A)=\boldsymbol{\alpha}_A.\epsilon_0\boldsymbol{G}(\boldsymbol{r}_A,\boldsymbol{ r}_D,\omega).\boldsymbol{p}_D
\end{eqnarray}
allowing us to write the normalized FRET transfer as
\begin{eqnarray}
\frac{\Gamma_{DA}}{\Gamma_0}= \frac{\epsilon_0 Im[\alpha_A(\omega)]|\boldsymbol{\mu}_A.\boldsymbol{G}(\boldsymbol{r}_A,\boldsymbol{ r}_D,\omega).\boldsymbol{\mu}_D|^2}{Im[\boldsymbol{\mu}_D .\boldsymbol{G}_0(\boldsymbol{r}_D,\boldsymbol{ r}_D,\omega). \boldsymbol{\mu}_D]}
\end{eqnarray}
and using the absorption cross section expression $\sigma_A(\omega)=\frac{\omega}{3c} Im[\alpha_A(\omega)]$ and the imaginary part of the free space dyadic green function $Im(\boldsymbol{G}_0)=\frac{k^3}{6\pi\epsilon_0}\boldsymbol{I}$
it leads to
\begin{eqnarray}
\frac{\Gamma_{DA}}{\Gamma_0}=18\pi \epsilon_0^2 c^4  \frac{\sigma_A(\omega)|\boldsymbol{\mu}_A.\boldsymbol{G}(\boldsymbol{r}_A,\boldsymbol{ r}_D,\omega).\boldsymbol{\mu}_D|^2}{\omega^4}\text{d} \omega.
\end{eqnarray}
Introducing  the normalized emission spectrum of the donor $f_D(\omega) $, we finally obtain the important result
\begin{eqnarray}
\frac{\Gamma_{DA}}{\Gamma_0}=18\pi \epsilon_0^2 c^4 \int_0^{\infty} \frac{\sigma_A(\omega)f_D(\omega)M(\omega)}{\omega^4}\text{d} \omega
\end{eqnarray}
with
\begin{eqnarray}
M(\omega)= |\boldsymbol{\mu}_A.\boldsymbol{G}(\boldsymbol{r}_A,\boldsymbol{ r}_D,\omega).\boldsymbol{\mu}_D|^2.
\end{eqnarray}

\begin{acknowledgments}
\section{ Acknowledgements}
This work has been supported by the EU Project {\it Nanomagna} under contract NMP3-SL-2008-214107.
\end{acknowledgments}

\end{document}